\documentclass[twocolumn,prl,showpacs,preprintnumbers,
               superscriptaddress,floatfix]{revtex4}

\usepackage{amssymb}
\usepackage{graphicx}
\usepackage{dcolumn}
\usepackage{bm}
\usepackage{longtable}
\begin{document}

\title{First order shape transition and critical point nuclei in Sm isotopes
       from relativistic mean field approach}

\author{J. Meng}
 \email{mengj@pku.edu.cn}
 \affiliation{School of Physics, Peking University, Beijing 100871, China}
 \affiliation{Institute of Theoretical Physics, Chinese Academy of Sciences,
              Beijing 100080, China}
 \affiliation{Center of Theoretical Nuclear Physics, National Laboratory of
              Heavy Ion Accelerator, Lanzhou 730000, China}
\author{W.  Zhang}
 \affiliation{School of Physics, Peking University, Beijing 100871, China}
\author{S.-G. Zhou}
 \affiliation{School of Physics, Peking University, Beijing 100871, China}
 \affiliation{Max-Planck-Institut f\"ur Kernphysik, 69029 Heidelberg, Germany}
 \affiliation{Institute of Theoretical Physics, Chinese Academy of Sciences,
              Beijing 100080, China}
 \affiliation{Center of Theoretical Nuclear Physics, National Laboratory of
              Heavy Ion Accelerator, Lanzhou 730000, China}
\author{H. Toki}
 \affiliation{Research Center for Nuclear Physics (RCNP), Osaka University,
              Ibaraki, Osaka 567-0047, Japan}
\author{L.S. Geng}
 \affiliation{School of Physics, Peking University, Beijing 100871, China}
 \affiliation{Research Center for Nuclear Physics (RCNP), Osaka University,
              Ibaraki, Osaka 567-0047, Japan}

\date{\today}

\begin{abstract}
The critical point nuclei in Sm isotopes, which marks the first
order phase transition between spherical U(5) and axially deformed
shapes SU(3), have been investigated in the microscopic quadrupole
constrained relativistic mean field (RMF) model plus BCS method
with all the most used interactions, i.e., NL1, NL3, NLSH and TM1.
The calculated potential energy surfaces show a clear shape
transition for the even-even Sm isotopes with $N = 82\sim 96$ and
the critical point nuclei are found to be $^{148}$Sm, $^{150}$Sm
and $^{152}$Sm. Similar conclusions can also be drawn from the
microscopic neutron and proton single particle spectra.
\end{abstract}


\maketitle

The equilibrium shape for finite many-body systems such as
nucleus, atom, molecule, etc. have been a hot topic for the past
several decades. These shapes are normally not rigid and change as
a function of their constituents. Sometimes these changes can be
quite abrupt and exhibit phase transitional and critical point
behavior similar to that found in a wide variety of many-body
systems, though the concepts of phase transitions are only
approximate for finite system.

Properties of the system in the transition region and, in
particular, at the critical point can be found by solving
numerically the eigenvalue problem for the Hamiltonian. However,
the transition regions are difficult to interpret as they exhibit
a complicated interplay of competing degrees of freedom. Yet such
system are in many respects the most important, as their structure
defines the nature of the transition region itself.

Recently, the concept of such critical point solutions has been
introduced in the framework of the algebraic models, in which
different shapes (phases) correspond to dynamic symmetries of some
algebraic structure \textit{G} and the phase transition
corresponds to the breaking of the dynamical symmetries. In the
Interacting Boson Model~\cite{iacbook} \textit{G} $\equiv$ U(6)
and there are three dynamical symmetries characterized by the
first algebra in the chain originating from U(6), that is U(5),
SU(3), and SO(6), with spherical, axially deformed, and
$\gamma$-unstable shapes. Experimental examples of all three types
of symmetries can be found in nuclei, e.g., the U(5) nuclei can
occur near closed shells, SU(3) nuclei can occur in the middle of
shells, and the O(6) limit tends to occur in nuclei with
particles-holes configuration. The phase transition between
spherical U(5) and axially deformed shapes SU(3) is first order
\cite{diep,feng}. The phase/shape transition region is
characterized by a pronounced $\beta$ softness~\cite{iac98}. This
characteristic led to the analytic solution for critical point
nuclei, called X(5) for axially symmetric case, and is based on
analytic solutions of the Schr\"{o}dinger equation corresponding
to a geometric (Bohr) Hamiltonian with a square-well
potential~\cite{iac01}. It has been shown that
$^{152}$Sm~\cite{cas01} and other $N$ = 90 isotones~\cite{cap02}
are the first empirical manifestation of the predicted nuclei at
the critical point of vibrator (spherical shape) to axial rotor
(axially deformed shape) phase transition.

The critical-point symmetries provide a classification of states
and analytic expressions for observables in regions where the
structure changes most rapidly. The ideas can be visualized in
terms of two coexisting potentials, spherical and deformed, whose
energy separation varies with nucleon number. Although phase
transitions and critical point have been discussed in atomic
nuclei for many years, there is no microscopical investigations
yet. As an important complementary way, a microscopic study could
be very helpful for us to understand where and how nuclei
exhibiting structures close to these symmetries occur. It is the
main purpose of this Letter to close this gap in the study of
symmetry.  The relativistic mean field (RMF) theory
~\cite{Serot:1986} has received wide attention because of its
successful description of many nuclear phenomena during the past
years~\cite{Ring:1996}. In particular, the exotic halo phenomena
can be understood self-consistently in this microscopical model
after the proper treatment of the continuum
effect~\cite{Meng:1996,Meng:1998a,Meng:1998}.  Here in this letter
we will discuss the realization of the critical point concept in
atomic nuclei in the microscopic quadrupole constrained RMF theory
with pairing treated by the BCS method, which will show that
$^{148,150,152}$Sm are excellent empirical manifestations of this
critical point structure.

The basic Ansatz of the RMF theory is a Lagrangian density where
nucleons are described as Dirac particles which interact via the
exchange of various mesons including the isoscalar-scalar $\sigma$
meson, the isoscalar-vector $\omega$ meson and the
isovector-vector $\rho$ meson. The effective Lagrangian density
considered is written in the form:
\begin{eqnarray}
\displaystyle
 {\cal L}
   & = &
     \bar\psi_i \left( i\rlap{/}\partial -M \right) \psi_i
    + \frac{1}{2} \partial_\mu \sigma \partial^\mu \sigma
    - U(\sigma)
    - g_{\sigma} \bar\psi_i \sigma \psi_i
   \nonumber \\
   &   & \mbox{}
    - \frac{1}{4} \Omega_{\mu\nu} \Omega^{\mu\nu}
    + \frac{1}{2} m_\omega^2 \omega_\mu \omega^\mu
    - g_{\omega} \bar\psi_i \rlap{/}{\mbox{\boldmath$\omega$}} \psi_i
   \nonumber \\
   &   & \mbox{}
    - \frac{1}{4} \vec{R}_{\mu\nu} \vec{R}^{\mu\nu}
    + \frac{1}{2} m_{\rho}^{2} \vec{\rho}_\mu \vec{\rho}^\mu
    - g_{\rho} \bar\psi_i \rlap{/} \vec{{\mbox{\boldmath$\rho$}}} \vec{\tau} \psi_i
   \nonumber \\
   &   &\mbox{}
    - \frac{1}{4} F_{\mu\nu} F^{\mu\nu}
    - e \bar\psi_i \frac{1-\tau_3}{2}\rlap{/}{\bf A} \psi_i ,
 \label{lagrangian}
\end{eqnarray}
where $\bar\psi=\psi^\dag\gamma^0$ and $\psi$ is the Dirac spinor.
Other symbols have their usual meanings.

\begin{table}
\caption{\label{table:binding} The total binding energy and the
quadrupole deformation $\beta_2$ of the ground-states of
$^{144-158}$Sm calculated by the constrained RMF theory with the
parameter sets NL1, NL3, NLSH and TM1.}
\begin{ruledtabular}
\begin{tabular}{c|rrrrr}
E[MeV]      & EXP\textsuperscript{\cite{Audi:1997}} & NL1 & NL3 & NLSH & TM1 \\
\hline\hline
%            & \multicolumn{5}{c}{Total binding energy [MeV]} \\
%\hline
 $^{144}$Sm &   1195.74&   1201.28&   1198.34&   1200.20&      1201.63\\
 $^{146}$Sm &   1210.91&   1213.83&   1211.79&   1214.00&      1215.89\\
 $^{148}$Sm &   1225.40&   1225.95&   1225.48&   1227.95&      1230.12\\
 $^{150}$Sm &   1239.25&   1239.59&   1239.29&   1241.09&      1243.54\\
 $^{152}$Sm &   1253.11&   1252.78&   1253.54&   1255.39&      1256.12\\
 $^{154}$Sm &   1266.94&   1265.50&   1267.54&   1269.59&      1268.99\\
 $^{156}$Sm &   1279.99&   1276.98&   1280.05&   1282.26&      1281.61\\
 $^{158}$Sm &   1291.98&   1287.57&   1291.73&   1293.99&      1293.07\\
\hline\hline
$\beta_2$   & EXP\textsuperscript{\cite{MNM:1995}} & NL1 & NL3 & NLSH & TM1 \\
%          & \multicolumn{5}{c}{Quadrupole deformation $\beta_2$} \\
\hline
 $^{144}$Sm &      0.00&      0.00&      0.00&   $-$0.01&      0.00\\
 $^{146}$Sm &      0.00&      0.05&      0.06&      0.08&      0.06\\
 $^{148}$Sm &      0.16&      0.15&      0.14&      0.14&      0.13\\
 $^{150}$Sm &      0.21&      0.25&      0.23&      0.21&      0.16\\
 $^{152}$Sm &      0.24&      0.34&      0.30&      0.28&      0.19\\
 $^{154}$Sm &      0.27&      0.35&      0.33&      0.32&      0.31\\
 $^{156}$Sm &      0.28&      0.36&      0.34&      0.33&      0.33\\
 $^{158}$Sm &      0.28&      0.36&      0.35&      0.34&      0.34\\
\end{tabular}
\end{ruledtabular}
\end{table}

The Dirac equation for the nucleons and the Klein-Gordon type
equations for the mesons and the photon are given by the
variational principle and can be solved by expanding the
wavefunctions in terms of the eigenfunctions of a deformed axially
symmetric harmonic-oscillator potential \cite{Gambhir:1990} or a
Woods-Saxon potential\cite{ZMR03}. The details can be also found
in Ref.~\cite{Ring:1996} and references therein.

The potential energy curve can be calculated microscopically by
the constrained RMF theory. The binding energy at certain
deformation value is obtained by constraining the quadrupole
moment $\langle Q_2 \rangle$ to a given value $\mu_2$ in the
expectation value of the Hamiltonian~\cite{Ring:1980},
\begin{equation}
  \langle H'\rangle~
   =~\langle H\rangle
    +\displaystyle\frac{1}{2} C_{\mu} \left(\langle Q_2\rangle -\mu_2\right)^2,
\end{equation}
where $C_{\mu}$ is the constraint multiplier.

For the nuclei studied in this paper, the full $N$ = 20 deformed
harmonic oscillator shells are taken into account and the
convergence of the numerical calculation on the binding energy and
the deformation is very good. The converged deformations
corresponding to different $\mu_2$ are not sensitive to the
deformation parameter $\beta_0$ of the harmonic oscillator basis
in a reasonable range due to the large basis. The different
choices of $\beta_0$ lead to different iteration numbers of the
self-consistent calculation and different computational time. But
physical quantities such as the binding energy and the deformation
change very little. Thus the deformation parameter $\beta_0$ of
the harmonic oscillator basis is chosen near the expected
deformation to obtain high accuracy and low computation time cost.
By varying $\mu_2$, the binding energy at different deformation
can be obtained. The pairing is considered by the constant gap
approximation (BCS) in which the pairing gap is taken as
$\textstyle\displaystyle\frac{12}{\sqrt{A}}$ for even number
nucleons.

\begin{figure}
\includegraphics[width=8.5cm]{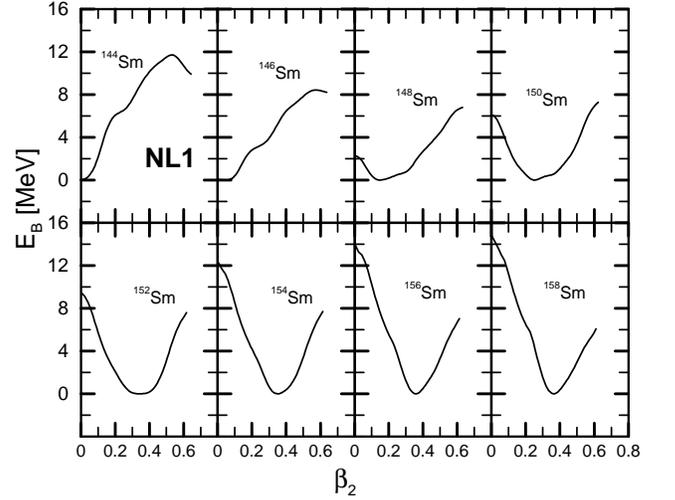}
\caption{The potential energy curves for $^{144-158}$Sm obtained
by the constrained RMF theory with the NL1 parameter set. The
ground state binding energy is taken as a reference. }
\label{fig:smnl1}
\end{figure}

\begin{figure}
\includegraphics[width=8.5cm]{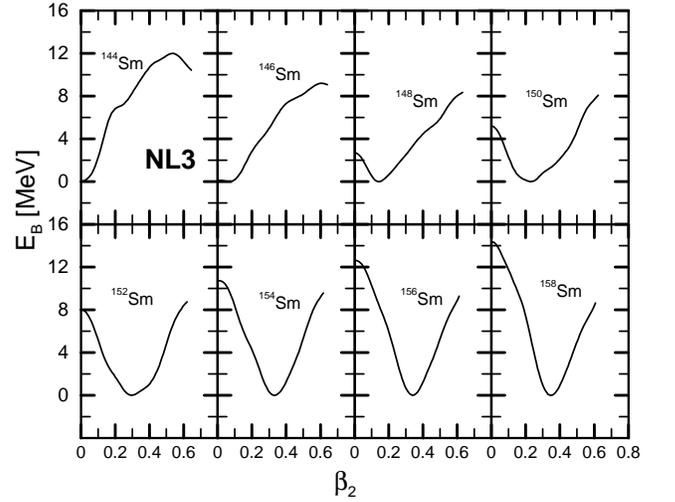}
\caption{The same as Fig.~\ref{fig:smnl1}, but with NL3}
\label{fig:smnl3}
\end{figure}

The binding energies and the quadropole deformation for the ground
states are listed in Table~\ref{table:binding} in constrained RMF
theory with NL1\cite{RRMG:1986}, NLSH\cite{SNR:1993},
TM1\cite{ST:1994} and NL3\cite{LKR:1997} parameter sets. For the
binding energies, the data are well reproduced within 0.5~\%.
Particularly for NL3, excellent agreement (within 1 MeV) is
observed for the binding energy in $^{146-158}$Sm. Even for
neutron magic nuclei $^{144}$Sm, the difference between the RMF
calculation and the data is less than 3 MeV, i.e., relatively less
than 0.3~\%. The spherical shapes in $^{144,146}$Sm and the weakly
deformed $^{148}$Sm are well reproduced. The deformations in
$^{150-158}$Sm are a little over-estimated by the theoretical
calculations. In general, it can be concluded that the data is
well reproduced by the constrained RMF theory.

\begin{figure}
\includegraphics[width=8.5cm]{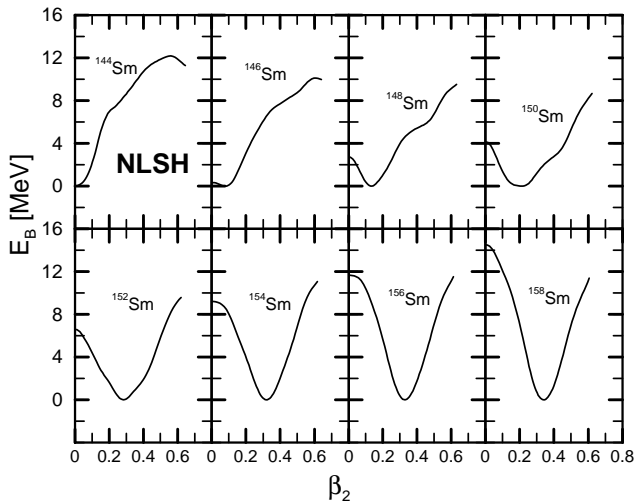}
\caption{The same as fig.~\ref{fig:smnl1}, but with NLSH}
\label{fig:smnlsh}
\end{figure}

\begin{figure}[tbp]
\includegraphics[width=8.5cm]{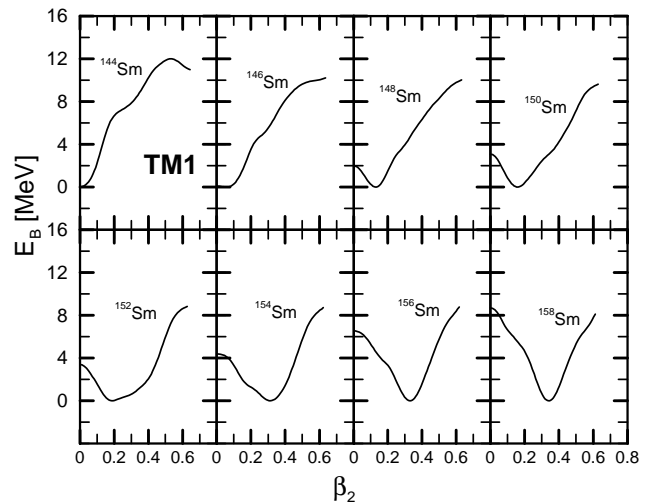}
\caption{The same as Fig.~\ref{fig:smnl1}, but with TM1}
\label{fig:smtm1}
\end{figure}

Figures ~ \ref{fig:smnl1}, \ref{fig:smnl3}, \ref{fig:smnlsh}, and
\ref{fig:smtm1} show the potential energy curves for
$^{144-158}$Sm in constrained RMF theory with NL1, NL3, NLSH and
TM1 parameters, in which the energy for the ground state is taken
as a reference. Similar patterns are found for all the parameter
sets. The ground state of $^{144}$Sm is found to be spherical and
has about 12 MeV stiff barrier against deformation. Although the
ground state of $^{146}$Sm is still spherical, its barrier becomes
around 8 MeV against deformation. With the increase of the neutron
number, the ground state gradually moves toward the deformed side
and the potential energy curve becomes more soft. Finally well
deformed $^{154-158}$Sm are observed.

\begin{table}
\caption{\label{table:bindingdifference} The difference of the
binding energy between the spherical state and the ground state in
unit of MeV calculated by the constrained RMF theory with the NL1,
NL3, NLSH and TM1 parameter sets for $^{144-158}$Sm. }
\begin{ruledtabular}
\begin{tabular}{c|rrrr}
            &    NL1 &    NL3 &   NLSH &    TM1 \\
\hline
 $^{144}$Sm &   0.00 &   0.00 &   0.00 &   0.00 \\
 $^{146}$Sm &   0.02 &   0.11 &   0.33 &   0.07 \\
 $^{148}$Sm &   2.26 &   2.67 &   2.72 &   1.98 \\
 $^{150}$Sm &   6.11 &   5.19 &   4.13 &   3.08 \\
 $^{152}$Sm &   9.51 &   8.02 &   6.59 &   3.38 \\
 $^{154}$Sm &  12.33 &  10.73 &   9.20 &   4.37 \\
 $^{156}$Sm &  14.10 &  12.61 &  11.64 &   6.53 \\
 $^{158}$Sm &  14.80 &  14.36 &  14.51 &   8.68 \\
\end{tabular}
\end{ruledtabular}
\end{table}

To examine how the shape of the Sm isotopes changes with the
neutron number, the differences of the binding energy between the
ground state and the state with spherical shape for Sm isotopes
are presented in Table~\ref{table:bindingdifference} in the
constrained RMF theory with the NL1, NL3, NLSH and TM1 parameter
sets. These differences can tell us how soft the nucleus is
against deformation and may give us a hint on the phase transition
of the nuclear shape. From $^{144-158}$Sm, the energy differences
between the ground state and the state with spherical shape change
from 0 to 15 MeV. There are two jumps in the energy differences.
The first jump appears at $^{148}$Sm and the second at $^{154}$Sm,
which suggests the shape transition from spherical to critical
point nuclei, and finally to axially deformed nuclei. The
potential energy curves for $^{148}$Sm, $^{150}$Sm, and $^{152}$Sm
are relatively flat, i.e., they are $\beta_2$-soft nuclei in the
transition between spherical and axially symmetric deformed
nuclei.

\begin{figure}
\includegraphics[width=8.5cm]{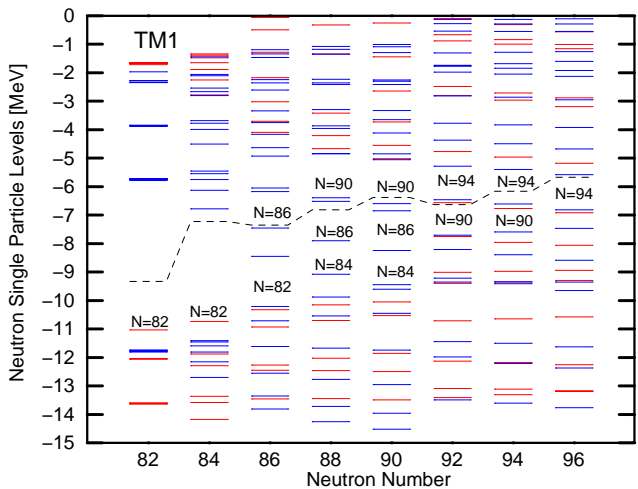}
\caption{The single neutron levels for $^{144-158}$Sm obtained by
the constrained RMF theory with the TM1 parameter set.}
\label{fig:smtm1nf}
\end{figure}

\begin{figure}
\includegraphics[width=8.5cm]{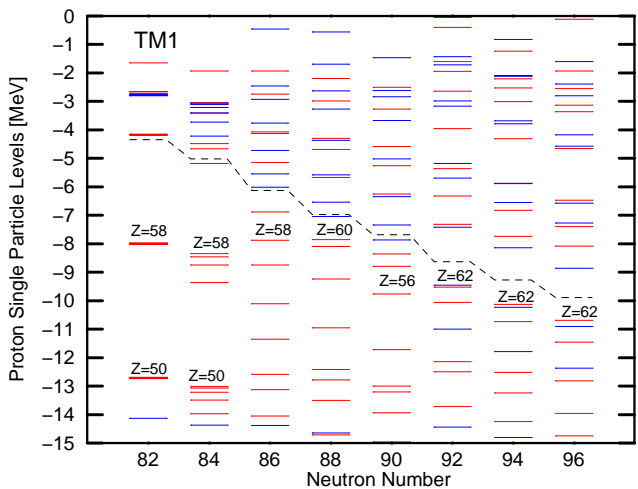}
\caption{The single proton levels for $^{144-158}$Sm obtained by
the constrained RMF theory with the TM1 parameter set.}
\label{fig:smtm1pf}
\end{figure}

One of the merits of microscopic nuclear models such as RMF theory
is that it can provide detailed information on single particle
levels, shell structure etc., which are very important for us to
discuss nuclear structure and examine the deformation driving
effect. In Fig. \ref{fig:smtm1nf} the single neutron levels for
$^{144-158}$Sm lying between 0 and $-$15 MeV are shown. The Fermi
levels are presented by dashed line. Fig. \ref{fig:smtm1pf} is
similar to Fig. \ref{fig:smtm1nf} but for protons. These two
figures present results calculated with the parameter set TM1. The
other three parameter sets give similar single particle structure
thus not presented here.

From Figs.~\ref{fig:smtm1nf} and \ref{fig:smtm1pf} one finds a
consistency between the shell structure evolution and the shape
evolution in Sm isotopes with increasing neutron number. Namely,
for $^{144}$Sm, spherical symmetry is better restored. The
deformation develops with $N$ increasing.  The deformation in
$^{146}$Sm is still small and the energy gap with $N=82$ can be
clearly seen from the single particle spectra. The nuclei
$^{148-150}$Sm  belong to a transition area, in which the $N=82$
gaps still exist but much smaller than that in $^{144,146}$Sm.
Starting from $^{154}$Sm, the gap with $N=82$ disappears.
Meanwhile a deformed gap for $Z=62$ develops and also that for
$N=94$.  Correspondingly, we observe the well deformed
$^{154-158}$Sm.

In summary, the shape transition between spherical and axially
deformed nuclei in $^{144-158}$Sm are investigated by the
microscopic quadropole constrained relativistic mean field theory
with all the most used interactions, i.e., NL1, NL3, NLSH and TM1.
The RMF calculation reproduces very well the data of the binding
energy and deformation for the ground states. By examining the
potential energy curves and the single particle levels obtained by
this microscopic approach, $^{148}$Sm ,$^{150}$Sm, and $^{152}$Sm
are found to be soft against $\beta$ deformation and the  critical
point candidate nuclei, which marks the first order phase
transition between spherical U(5) and axially deformed shapes
SU(3).

\begin{acknowledgments}
We thank Victor Zamfir for helpful discussions and useful
comments. This work was partly supported by the Major State Basic
Research Development Program Under Contract Number G2000077407 and
the National Natural Science Foundation of China under Grant No.
10025522, 10221003 and 10047001.
\end{acknowledgments}

\end{document}